\documentclass[preprint,showpacs,preprintnumbers,amsmath,amssymb]{revtex4}

\usepackage{graphicx}
\usepackage{dcolumn}
\usepackage{bm}

\begin{document}

\title{\large Observation of two Andreev-like energy scales in
$La_{2-x}Sr_{x}CuO_4$ superconductor/normal-metal/superconductor
junctions}

\author{G. Koren}
\email{gkoren@physics.technion.ac.il} \affiliation{Physics
Department, Technion - Israel Institute of Technology Haifa,
32000, ISRAEL} \homepage{http://physics.technion.ac.il/~gkoren}

\author{T. Kirzhner}
\affiliation{Physics Department, Technion - Israel Institute of
Technology Haifa, 32000, ISRAEL}

\date{\today}
\def\bfig {\begin{figure}[tbhp] \centering}
\def\efig {\end{figure}}

\normalsize \baselineskip=8mm  \vspace{15mm}


\footnotemark{****************************************************************\\
A slightly shortened version of this paper is published in \\
Phys. Rev. Lett. \textbf{106}, 017002 (2011)\\
http://link.aps.org/doi/10.1103/PhysRevLett.106.017002

\vspace{5mm}
\noindent This file includes also the supplementary material - starting on page 11 here\\
http://link.aps.org/supplemental/10.1103/PhysRevLett.106.017002
***********************************************************************}

\begin{abstract}

Conductance spectra measurements of highly transparent ramp-type
junctions made of superconducting $La_{2-x}Sr_{x}CuO_4$ electrodes
and non superconducting $La_{1.65}Sr_{0.35}CuO_4$ barrier are
reported. At low temperatures below $T_c$, these junctions have
two prominent Andreev-like conductance peaks with clear steps at
energies $\Delta_1$ and $\Delta_2$ with $\Delta_2 > 2\Delta_1$. No
such peaks appear above $T_c$. The doping dependence at 2 K shows
that both $\Delta_1$ and $\Delta_2$ scale roughly as $T_c$.
$\Delta_1$ is identified as the superconducting energy gap, while
a few scenarios are proposed as for the origin of $\Delta_2$.\\

\end{abstract}

\maketitle

The issue of two distinct energy gaps in the cuprates has been
discussed by many authors, and the question whether both are
related to superconductivity is still controversial
\cite{Renner,GDnature,Shen,Yoshida}. In one scenario, one energy
gap is the coherence gap which opens at $T_c$ with the onset of
phase coherent superconductivity, while the other gap opens at
$T^*$ which marks the cross over to the pseudogap regime and
possibly the creation of uncorrelated pairs \cite{Kivelson}. In
contrast to this scenario, some ARPES measurements show only a
single energy gap, which indicate that the superconducting gap and
the pseudogap might be the same entity \cite{Kanigel,Campuzano}.
In another scenario, the regime above $T_c$ in the underdoped
cuprates which exhibits a signature of the condensate, can be
attributed to strong superconducting fluctuations. This behavior
was found in measurements of the Nernst effect \cite{Ong}, whose
$T_c$(onset) values scale with doping roughly as the
superconducting dome. This effect therefore, is related to $T_c$
and apparently depends on more than one energy scale of the
condensate. Previous point contact measurements of tunneling and
Andreev conductance have shown that the tunneling gap which scales
as $T^*$ is larger than the Andreev gap which follows $T_c$
\cite{GDnature,GDreview,Achsaf,Gonnelli}. In the present study we
report on similar conductance measurements in ramp-type junctions
of the $La_{2-x}Sr_{x}CuO_4$ (LSCOx or LSCO) system, but due to
their high transparency we observe mostly Andreev gaps.
Surprisingly, we find two different such gaps in this system below
$T_c$ both of which scale versus doping roughly as the
superconducting dome. Only single gaps were observed in previous
conductance measurements in LSCOx \cite{Achsaf,Gonnelli,Yuli,Oda}.
The results though show that in Refs. \cite{Achsaf,Gonnelli,Yuli}
the gaps follow $T_c$ while in Ref. \cite{Oda} the gaps scale as
$T^*$. The present low energy Andreev peak in the conductance
spectra is attributed to the superconducting gap, while a few
scenarios are discussed in relation to the origin of the second
high energy feature in the spectra.\\

\begin{figure} \hspace{-20mm}
\includegraphics[height=9cm,width=13cm]{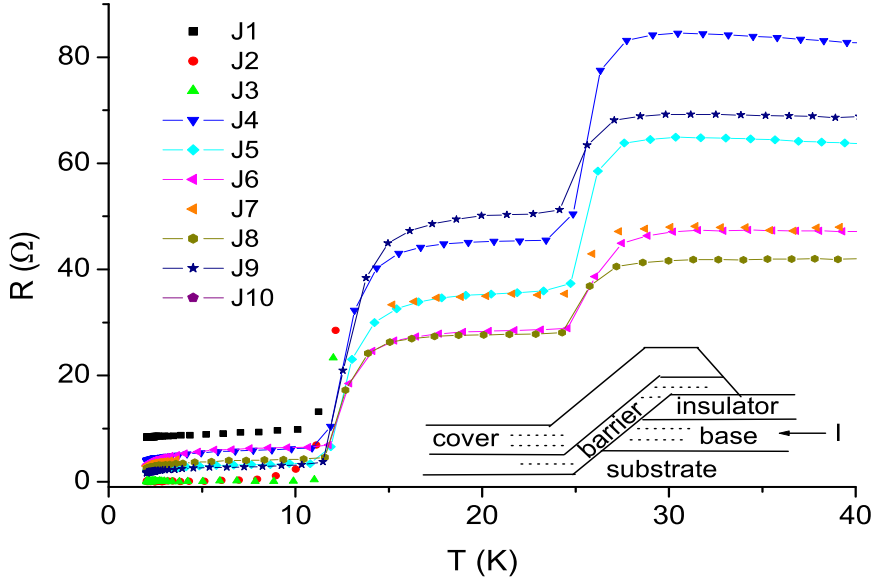}
\vspace{-0mm} \caption{\label{fig:epsart}Resistance versus
temperature of all the LSCO10 junctions on the wafer. The inset
shows a schematic drawing of a ramp-type junction, where the 77 nm
thick base and cover electrodes are made of LSCOx and the 33 nm
thick barrier is made of LSCO35.}
\end{figure}

Highly transparent superconductor - normal metal - superconductor
(SNS) junctions of the cuprates can be obtained if the S
electrodes and the N barrier have similar density of states and
Fermi velocities. In the LSCOx system the doping levels are
determined mostly by the Sr content, provided the same oxygen
annealing process is used. Therefore, highly transparent junctions
can be realized, if the S electrodes are in the superconducting
regime ($0.06 \leq x \leq 0.25$) while the N barrier is
non-superconducting with $x\approx 0.35-0.45$. This scenario
however, can not be realized in the $YBa_2Cu_3O_{7-\delta}$ (YBCO)
system, since one can not dope the S and N electrodes differently
with two different oxygen contents in the same junction. Pr and Fe
doped YBCO had been used in the past as barriers in SNS junctions
\cite{Emil,Nesher12,Koren}, but these dopants introduce larger
disturbances in the YBCO matrix as compared the different Sr
doping levels in the LSCO electrodes. We thus investigated
LSCOx-LSCO35-LSCOx ramp-type junctions with $x$ values of 0.1,
0.125, 0.15 and 0.18, in order to determine the various
Andreev-like gaps and study the doping dependence (or phase
diagram) of these gaps. Ten junctions were prepared for each
doping level along the anti-node direction in the geometry shown
in the inset to Fig. 1, on $1\times 1$ cm$^2$ wafers of (100)
$SrTiO_3$ (STO). All the different LSCO layers were grown
epitaxially with the c-axis normal to the wafer, and thus a-b
plane coupling was obtained between the base and cover electrodes.
All junctions had the same geometry with $5\,\mu$ width, and 77
and 33 nm films and barrier thicknesses, respectively. Typical
4-probe results of the resistance versus temperature for $x=0.1$
are shown in Fig. 1. One can easily see the two distinct
transition temperature onsets at 28 and 15 K, which correspond to
the $T_c$ values of the cover and base electrodes, respectively.
The reason for this is that the base electrode on the pristine STO
surface is more strained than the cover electrode which is grown
on a 33 nm thick LSCO35 layer on top of the ion milled area of the
STO wafer \cite{Locquet}. Below about 10 K, the quite constant
junctions resistance can be seen which ranges between
$0-8\,\Omega$ while most junctions have $2-4\,\Omega$. Junction 10
had bad contacts while junctions 2 and 3 have critical current.\\

\begin{figure} \hspace{-20mm}
\includegraphics[height=9cm,width=13cm]{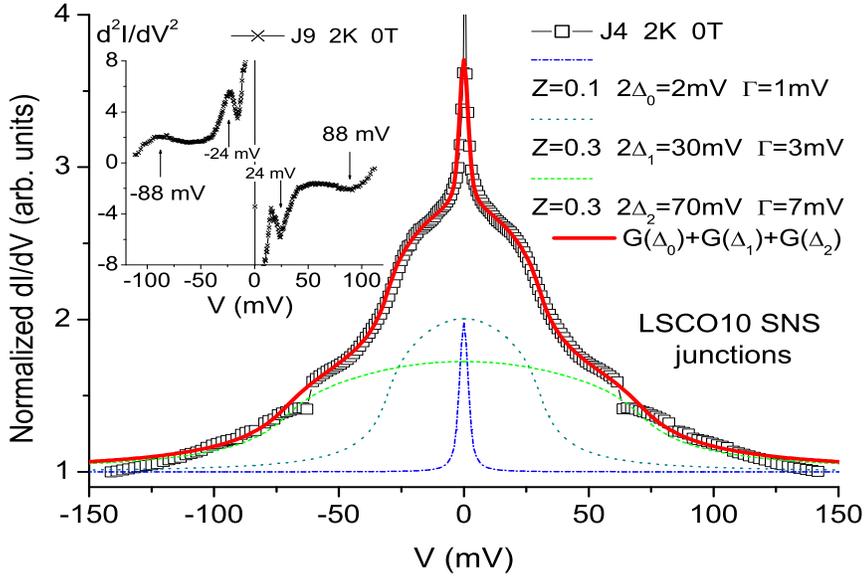}
\vspace{-0mm} \caption{\label{fig:epsart}Conductance spectrum of
an anti-node SNS junction of LSCO10-LSCO35-LSCO10 at 2K with a fit
to the BTK model for a d-wave superconductor. The three components
of the fit with $\Delta_0$, $\Delta_1$ and $\Delta_2$ are also
shown. The inset shows the derivative of the conductance data of
another junction on the same wafer. }
\end{figure}

Fig. 2 shows a representative normalized conductance spectrum at 2
K of the junction J4 of Fig. 1. This spectrum has three pronounced
features. The first is a narrow zero bias conductance peak (ZBCP),
the second is a dome like peak of intermediate width which is
superimposed on a third feature which is even broader. The
conductance data is therefore the result of a sum of three
components which can be written as
$G$(total)=$G(\Delta_0)+G(\Delta_1)+G(\Delta_2)$. Note that the
gap feature in SNS junction always appears at $2\Delta$
\cite{Tinkham}. Interference phenomena such as Tomasch
\cite{Tomasch} or McMillan-Rowel \cite{MMR} oscillations do not
affect this gap voltage and are absent in the present study
(apparently due to the very thick barrier and small ramp angle),
though they had been observed previously in similar YBCO based SNS
junctions \cite{Nesher12,Nesher3}. Furthermore, the use of
interference free SN junctions with a single interface involves
other problems in the determination of the voltage drop on the
junctions due to the much large voltage drop of the leads
\cite{SM}. We therefore decided to work with SNS junctions with
possible interference effects but with zero lead resistance and
accurate energy or voltage scale. We used the BTK model modified
for a d-wave superconductor given by Tanaka and Kashiwaya to fit
our data \cite{Tanaka}. The three conductance components
$G(\Delta_i)$ of these fits are shown in Fig. 2 together with the
total conductance curve $G$(total) which fits the data quite well.
The barrier strength $\rm Z_i$, the Andreev gap parameters
$\Delta_i$ and the lifetime broadening $\Gamma_i$ are also given
in Fig. 2. One can see that the $\rm Z_i$ values are quite low
which indicates a highly transparent junction. This justifies our
use of the anti-node direction formula without mixing of the node
direction, since both are quite similar when the $\rm Z_i$ values
are small. We also note that the maximum conductance value of each
component in Fig. 2 is at around 2 which is like the expected
Andreev value of the conductance of a pair for each incident
electron. Although this fitting procedure involves many
parameters, the clear Andreev-like gap features at $\Delta_1$ and
$\Delta_2$ can be deduced from the raw data directly by taking the
derivative of the conductance as shown in the inset. This was done
for a different junction on the same wafer, and one can see that
the peak locations are quite close to the different $2\Delta_i$
obtained before, but this also reflects the spread of these values
on the same wafer. Additional conductance spectra that show the
spread of the $2\Delta_i$ values are shown in Figs. 4S, 5S and 6S
of the supplementary material for LCO15-LSCO35-LSCO15 junctions
\cite{SM}. Fig. 3S there shows that the conductance spectra of
LSCO10-LSCO35 SN junctions \cite{SM} are basically quite similar
to the results of Fig. 2 here on SNS junctions. We note in passing
that the sharp resonances at $\pm 62$ mV in Fig. 2 are not very
common and appear in about one out of ten junctions on a wafer.\\

\begin{figure} \hspace{-20mm}
\includegraphics[height=9cm,width=13cm]{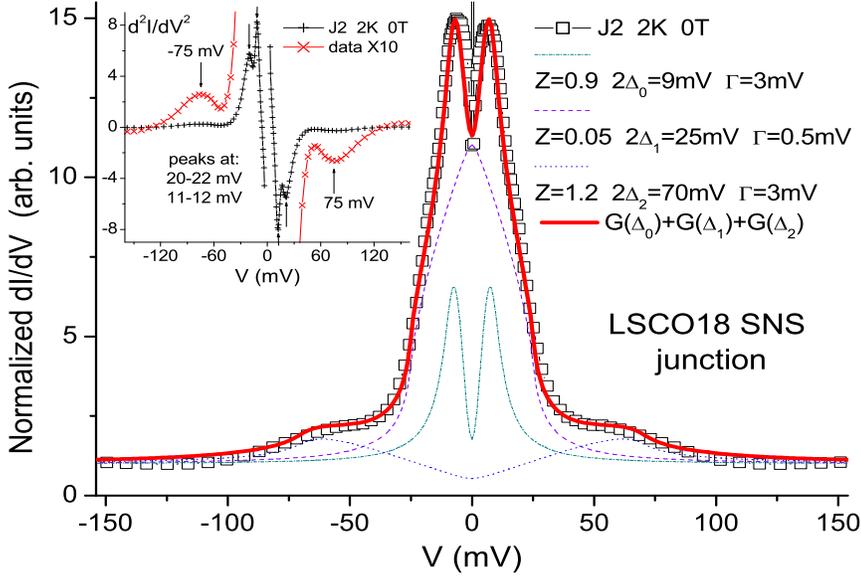}
\vspace{-0mm} \caption{\label{fig:epsart}Conductance spectrum of
an anti-node junction of LSCO18-LSCO35-LSCO18 at 2K with a fit to
the d-wave BTK model together with the three components of this
fit with $\Delta_0$, $\Delta_1$ and $\Delta_2$. The inset shows
the derivative of the conductance data of the main panel.}
\end{figure}

\begin{figure} \hspace{-20mm}
\includegraphics[height=17cm,width=13cm]{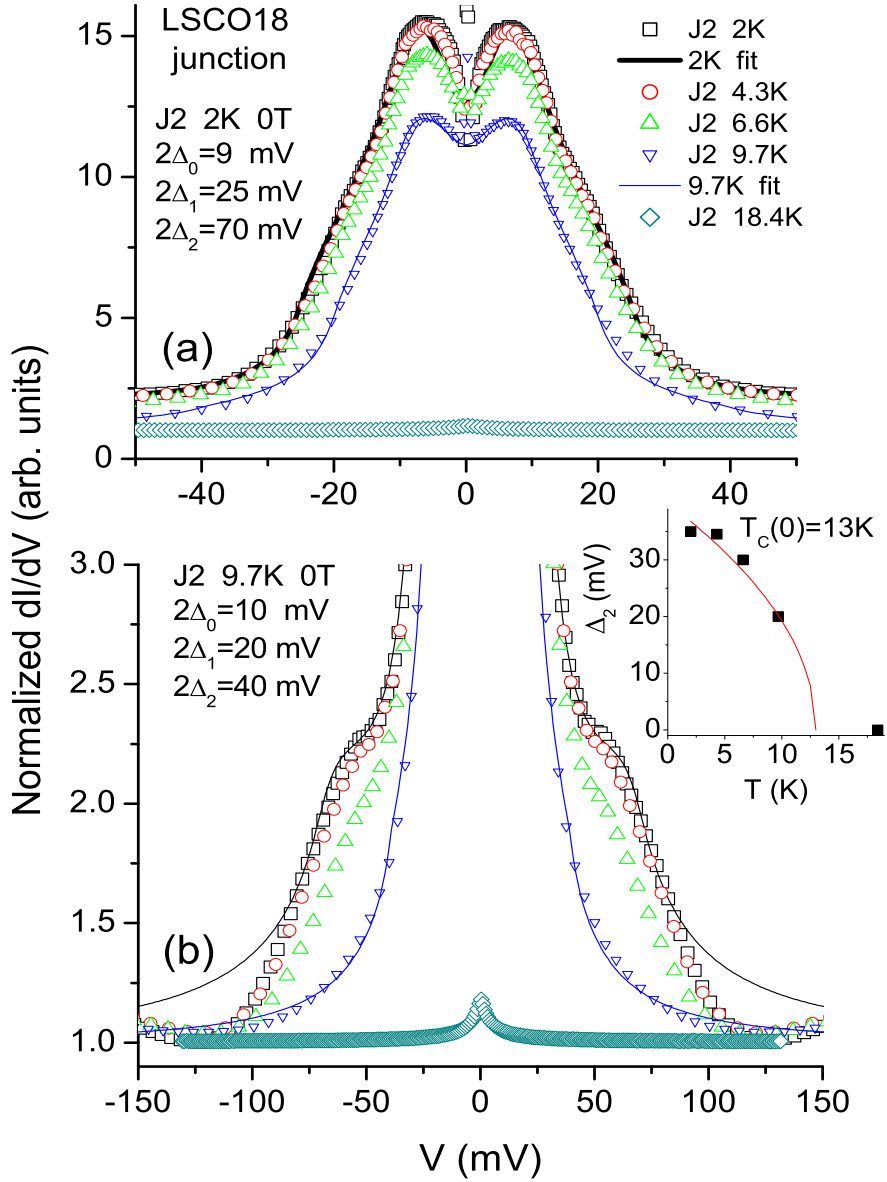}
\vspace{-0mm} \caption{\label{fig:epsart} Conductance spectra of
the same junction as in Fig. 3 at various temperatures T at low
bias 4(a), and up to high bias with zooming up on low conductances
4(b). The inset to 4(b) shows the large gap $\Delta_2$ behavior
versus T (squares) with a $\Delta_2(0)\sqrt{(T_c-T)/T_c}$ fit
(line).}
\end{figure}

A typical conductance spectrum of a LSCO18-LSCO35-LSCO18 junction
at 2 K together with a fit with its three components as before are
shown in Fig. 3. The dominant component contributing to this
spectrum is the highly transparent one at $\Delta_1$, but unlike
in Fig. 2, its maximum value now is above 10 and not around 2. We
attribute this behavior to the presence of bound states which can
cause this effect \cite{Tanaka}. The $\Delta_2$ feature is still
quite clear but has a small spectral weight as compared to that of
$\Delta_1$. It also has a lower transparency and shows a
tunneling-like behavior. The third feature near zero bias now
looks like a split ZBCP, again with intermediate transparency and
tunneling-like behavior. The very narrow ZBCP of Fig. 2 is gone,
and only a remnant critical current is observed. The inset to Fig.
3 shows the derivative of the conductance $d^2I/dV^2$ of the same
junction. One can see that the peak energies now are even closer
to those obtained from the fitting procedure in comparison to the
results of Fig. 2. Fig. 4 shows a few conductance spectra of the
same junction at different temperatures. As expected, both
$\Delta_1$ and $\Delta_2$ are suppressed with increasing
temperature while $\Delta_0$ is basically unaffected. The inset to
Fig. 4 (b) shows that $\Delta_2(T)$ behaves quite similarly to a
BCS gap versus temperature, and therefore can be considered as a
gap-like feature in the density of states. In addition, we found
that in all junctions above $T_c$ of both electrodes at about 30
K, all the conductance spectra were flat (not shown), which
indicates that no Andreev scattering could be observed. This is in
agreement with previous finding by Dagan \textit{et al.} in NIS
junctions \cite{Dagan}. Above $T_c$ however, the junction
contribution to the conductance is quite small compared to the
significant leads resistance, and any change due to possible
pairing in the pseudogap regime might be too small to be observed.
Conductance spectra were also measured under magnetic fields of up
to 6 T (not shown), and already at 2 T a strong suppression of all
the gap-like features was observed. We thus conclude that both
$\Delta_1$ and $\Delta_2$ represent gap-like features of the
LSCOx system.\\

\begin{figure} \hspace{-20mm}
\includegraphics[height=9cm,width=13cm]{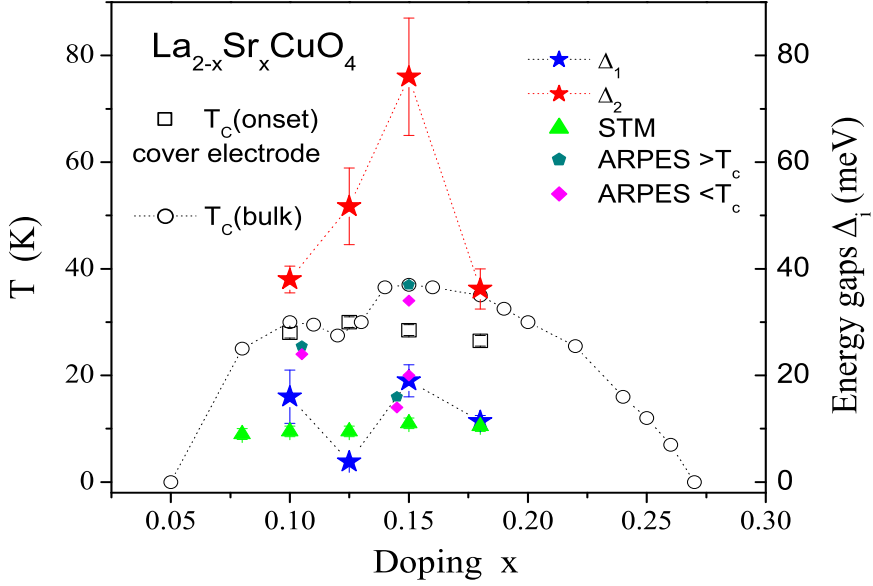}
\vspace{-0mm} \caption{\label{fig:epsart} The phase diagram of all
the LSCOx junctions versus doping $x$. Shown are the bulk and
cover electrode film transition temperatures, the two Andreev-like
energy gaps $\Delta_1$ and $\Delta_2$ of the present study at 2 K,
and previous STM gaps at 4.2 K \cite{Yuli} and ARPES gaps
\cite{Yoshida,Therashima,Shi}.}
\end{figure}

Fig. 5 summarizes on the phase diagram of LSCO the $\Delta_1$ and
$\Delta_2$ results of the present study at 2 K versus doping. Also
shown are STM and ARPES gaps \cite{Yuli,Yoshida,Shi,Therashima},
and the $T_c$ values of film and bulk LSCO \cite{bulk}. The
$\Delta_i$ values represent mean values of all working junctions
on the wafer for each doping level and their statistical error.
One can see that the general doping dependence of both $\Delta_1$
and $\Delta_2$ follows roughly the superconducting dome. The
$\Delta_2$ value at optimal doping of $x=0.15$ is strongly
enhanced by a factor of about two compared  the $\Delta_2$ values
at the 10\% and 18\% doping levels. This yields a peak-like
dependence on doping of $\Delta_2(x)$ rather than a dome. The
$\Delta_1$ value is strongly suppressed at the $x=1/8$ doping
level, similar to $T_c$. The $\Delta_1$ results agree with the STM
observations \cite{Yuli}, while the previous point contact results
with $\Delta\approx 6-8$ meV \cite{GDreview,Achsaf,Gonnelli} are
found on the lower side of the $\Delta_1$ values. Different ARPES
gaps for LSCOx were found by different groups at $x$=0.145 and
0.15 doping levels. Shi \textit{et al.} have measured $\Delta$=14
and 16 mV well below and well above $T_c$, respectively
\cite{Shi}, while the corresponding gaps that Therashima
\textit{et al.} have measured were $\Delta$=34 and 37 mV. The
former agree with our $\Delta_1$=19$\pm$3 mV value at $x$=0.15
which also agrees with Yoshida \textit{et al.} who measured
$\Delta_0 \approx 20$ meV \cite{Yoshida}, but the latter as well
as the ARPES gap of about 25 mV at $x$=0.105, fall in between the
present $\Delta_1$ and $\Delta_2$ values. Our results thus seem to
suggest that $\Delta_1$ is the superconducting gap. Its low value
at 1/8 doping also supports this conclusion if stripes are taken
into account \cite{Kivelson,stripes}. $\Delta_2$ seems to be
related to $T_c$, since it roughly follows its doping dependence,
but its origin is not so straight forward and
different scenarios for it will be discussed next.\\

First, since the $\Delta_2$ feature in the conductance spectra is
quite small, it might be attributed to a background "step down" in
the highly transparent junctions due to any excitation mode with
energy $\hbar\omega$ which will appear at
$eV=\hbar\omega-\Delta_1$ as discussed by Kirtley \cite{Kirtley}.
This result was obtained using a theory of inelastic transport at
the junctions interface and the excitations by the tunneling or
Andreev process with $\hbar\omega$ can be due to holons, bosons,
phonons an so on \cite{GDreview,Anderson}. This gives symmetric
spectra in agreement with the present results, but the doping
dependence of $\Delta_2$ as seen in Fig. 5 implies that these
excitations have to be related to superconductivity and the way
they actually do needs further theoretical treatment. A second
scenario for the appearance of the $\Delta_2$ feature is that it
might be related to the Van Hove singularity (VHS) in the 2D LSCO
system. Using the tt'J model it was shown that in addition to the
coherence peaks at the gap energy $\Delta$, two new and symmetric
peaks appear at 2-3 times $\Delta$ in the conductance spectra due
to the VHS \cite{Fedro}. This agrees with the present symmetric
spectra and the values of $\Delta_2$. However, when a tt't"J model
was used \cite{Hoogenboom}, asymmetric spectra were obtained which
disagree with our results but nevertheless, the peak energies are
still of the order of $\Delta_2$. The doping dependence that
follows from these results shows a monotonous increase of the
energy due to the VHS feature with decreased doping, similar to
the doping dependence of the pseudogap. This is in clear
contradiction to our results, but in view of the fact that the
calculations involved were done in attempt to explain the
asymmetrical tunneling spectra of BSCO \cite{Hoogenboom,Wen,Levy},
one can not rule out that further theoretical analysis for LSCO
might yield different results. Finally, although we are puzzled by
the possible existence of a proper Andreev gap at such high
energies as $\Delta_2$, the reasonably good fits to our data using
the d-wave BTK model \cite{Tanaka}, might indicate that $\Delta_2$
is originated in such a gap in the density of states. To relate
this to superconductivity as observed in Fig. 5, would need some
bold speculation as for instance the existence of pairs with an
even larger condensation energy. In this scenario then, $\Delta_2$
will be related to $\Delta_1$, but their relation to $T_c$ will
involve different doping dependent functions that will have to
account  for the fact that $\Delta_1(x=0.15)/\Delta_1( x=0.1)\sim
1$ while $\Delta_2(x=0.15)/\Delta_2( x=0.1)\sim 2$. Clearly, a
thorough theoretical modelling as for the origin of $\Delta_2$ is
needed, and this might add to our understanding of
the high temperature superconductors.\\

In conclusion, two Andreev-like energy gaps have been observed in
the LSCOx cuprates, both of which scale roughly with $T_c$ versus
doping. $\Delta_1$ is identified as the superconducting energy
gap, while the origin of $\Delta_2$ which is also related to
superconductivity, is unclear at the present time and needs
further theoretical modelling.\\

{\em Acknowledgments:}  This research was supported in part by the
Israel Science Foundation (grant \# 1096/09), the joint
German-Israeli DIP project and the Karl Stoll Chair in advanced
materials at the Technion.\\






\newpage

\begin{center}
\Large Supplementary material for:\\
\vspace{3mm} Observation of two Andreev-like energy scales
in $La_{2-x}Sr_{x}CuO_4$\\
superconductor/normal-metal/superconductor junctions
\end{center}

\date{\today}
\def\bfig {\begin{figure}[tbhp]} 
\def\efig {\end{figure}}

\normalsize \baselineskip=6mm  \vspace{25mm}


\section{Preliminary studies of S-S and S-N
ramp-junctions} \normalsize \baselineskip=6mm  \vspace{6mm}

We started the LSCO based ramp-junctions project by preparing and
characterizing S-S "shorts" which are S-N-S junctions without the
barrier. This was done in order to test the quality and
cleanliness of our fabrication process and check the quality of
the contact at the junction interface. We measured the I-V curves
of LSCO10-LSCO10 shorts and extracted the critical current density
at 2 K ($J_c(2K)$) by the use of the 1 $\mu V$ criterion. We found
that the highest values of $J_c(2K)$ ranged between $3-5\times
10^6\,A/cm^2$. When compared with $J_c(2K)$ of similar LSCO10
microbridges which is typically of about $10\times 10^6\,A/cm^2$,
we can conclude that considering the complexity of the multi-step
fabrication process of the S-S shorts \cite{Nesher12}, their
quality is fairly good.\\

\begin{figure} \hspace{-20mm}
\includegraphics[height=9cm,width=13cm]{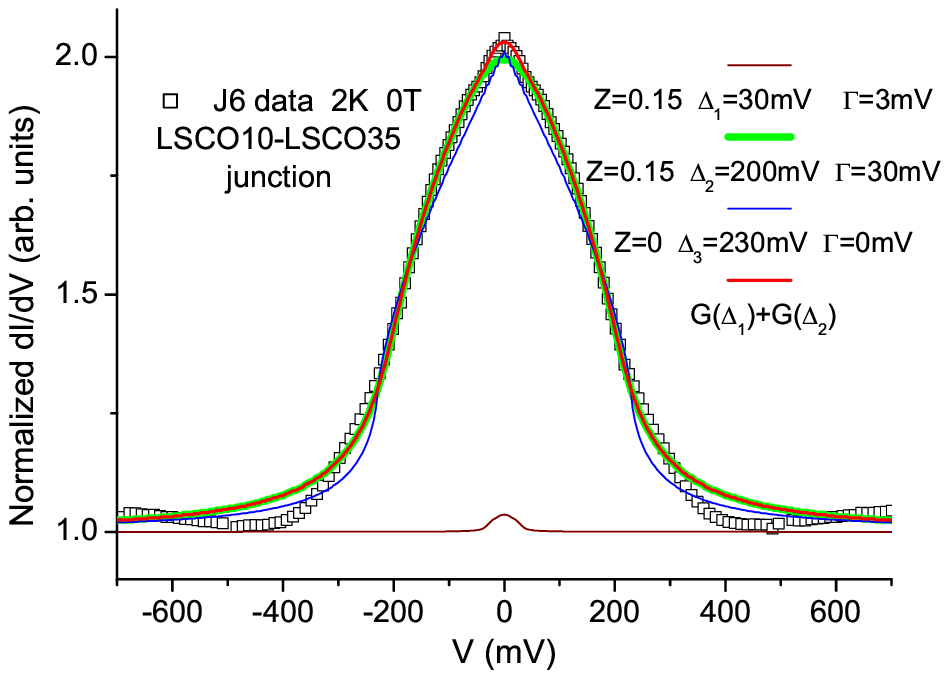}

\noindent {Fig. 1S:  Conductance spectrum of an anti-node S-N
junction of LSCO10-LSCO35 at 2K with a fit to the BTK model for a
d-wave superconductor. The two components of the fit with
$\Delta_1$ and $\Delta_2$ are shown together with a less good fit
with a single gap parameter $\Delta_3$.}
\end{figure}

\begin{figure} \hspace{-20mm}
\includegraphics[height=9cm,width=13cm]{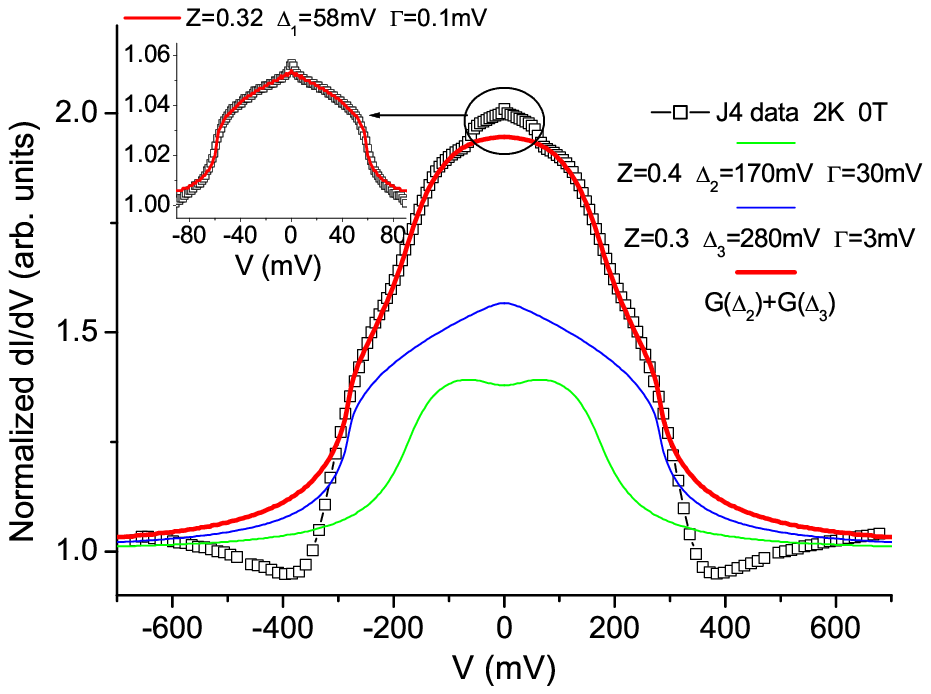}

\noindent Fig. 2S:  Conductance spectrum of another anti-node S-N
junction of LSCO10-LSCO35 on the same wafer at 2K with a fit to
the BTK model for a d-wave superconductor. The two components of
the fit with $\Delta_2$ and $\Delta_3$ are also shown. In the
inset, the low bias data in the marked circle of the main panel is
fitted with a single gap value of $\Delta_1$.
\end{figure}

Next we prepared S-N junctions for spectroscopic characterization.
We have chosen to work first with S-N junctions for two reasons.
One is that having a single interface rather than two as in S-N-S
junctions prevents many interference effects, and the other is
that the higher resistance of the device would reduce the current
density in the junction and reduce heating and non-linear effects
at high bias. First we measured the resistance versus temperature
of LSCO10-LSCO35 S-N junctions. We found that the low bias
resistance of these junctions is in the range of 65-70 $\Omega$,
which is an order of magnitude higher than that of the
corresponding S-N-S junctions (see Fig. 1 of the main paper). This
leads to an order of magnitude lower current densities for the
same bias (at high biases however, this changes by a factor of
about 2 as can be seen in section III). Figs. 1S and 2S show the
conductance spectra of two different LSCO10-LSCO35 S-N junctions
on the same wafer, together with fits to the BTK model for a
d-wave superconductor \cite {Tanaka}. Although very prominent
Andreev-like peaks are observed with good fits to the modified BTK
model, the resulting energy gap values $\Delta_i$ are obviously
much too large. Even the smallest $\Delta_1$ values of 30 and 58
mV are much too high compared to the superconducting energy gap of
LSCO10 which should be in the 10-15 mV range. The reason for these
unphysically large $\Delta_i$ values is that the the voltage scale
V=$V_{measured}$ in Figs. 1S and 2S is actually the sum of the
voltage drop $V_{lead}$ on the lead resistance $R_{lead}$ of the
normal LSCO35 cover electrodes (of a few mm long film from the
voltage contact to the junction) and the voltage drop on the
junction $V_{junction}$. To correct this problem, one has to plot
the conductance versus a different V scale which is
$V_{junction}=V_{measured}-V_{lead}$ as was done in Fig. 3S. Once
this calibration is done, the resulting energy gap values are
$\Delta_1=15$ mV and $\Delta_2=50$ mV, which agree fairly well
with the corresponding values of 12-15 and 35-44 mV obtained from
the conductance spectra of the S-N-S junctions (see Figs. 2 and 5
of the main paper).\\

\begin{figure} \hspace{-20mm}
\includegraphics[height=9cm,width=13cm]{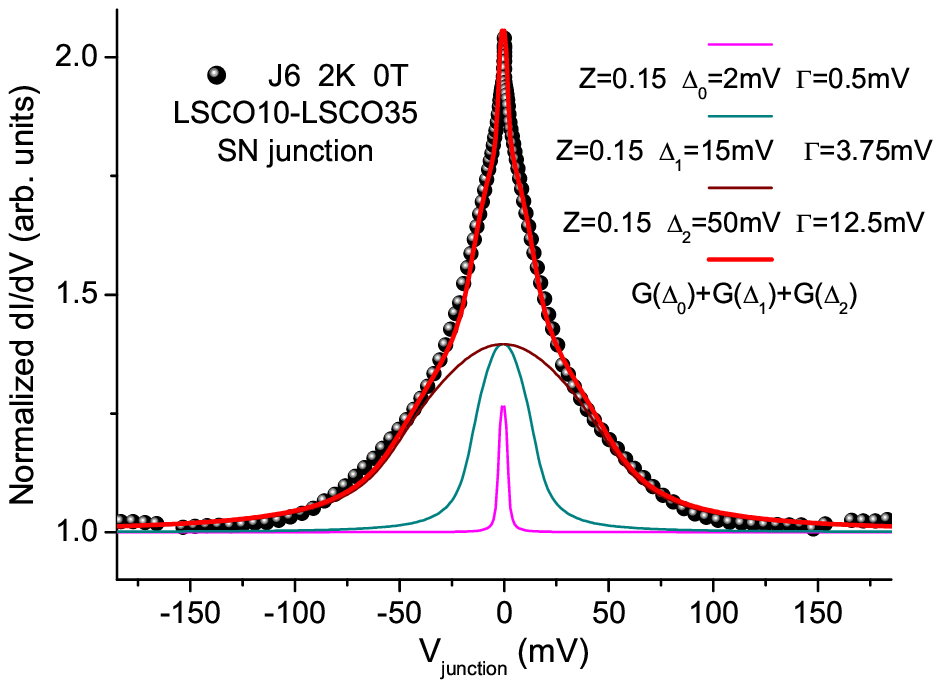}

\noindent Fig. 3S:  Conductance spectrum at 2K of the junction of
Fig. 1 with a calibrated V scale which is the net voltage drop on
the junction. A fit to the BTK model for a d-wave superconductor
is also shown with its three components $\Delta_0$, $\Delta_1$ and
$\Delta_2$.
\end{figure}

In order to perform the calibration procedure as noted above, we
integrated the conductance dI/dV data of Fig. 1S, and calibrated
the resulting $I-V_{measured}$ curve at low bias according to the
measured low bias resistance. Then the  values
$V_{lead}=R_{lead}I(V_{measured})$ were found, where $R_{lead}$
was calculated from the geometry of the leads and the resistivity
value of LSCO35 at 2K. This yielded the calibrated V scale
$V_{junction}=V_{measured}-V_{lead}$ of Fig. 3S. The problem with
this procedure is that due to the low junction resistance, this V
scale depends very sensitively on the subtraction of two similar
numbers $V_{measured}$ and $V_{lead}$, especially at low bias. Any
small deviation in the value of $V_{lead}$ due to slight thickness
or resistivity variations on different areas of the wafer would
lead to large deviations in $V_{junction}$, and sometimes even to
negative values. We therefore decided that the best way to get
reliable energy gap values is to use S-N-S junctions where the
lead resistance is zero as long as the electrodes are
superconducting. In this case, $V_{junction}=V_{measured}=V$ and
all the calibration problems of S-N junctions can be avoided.\\

\begin{figure} \hspace{-20mm}
\includegraphics[height=9cm,width=13cm]{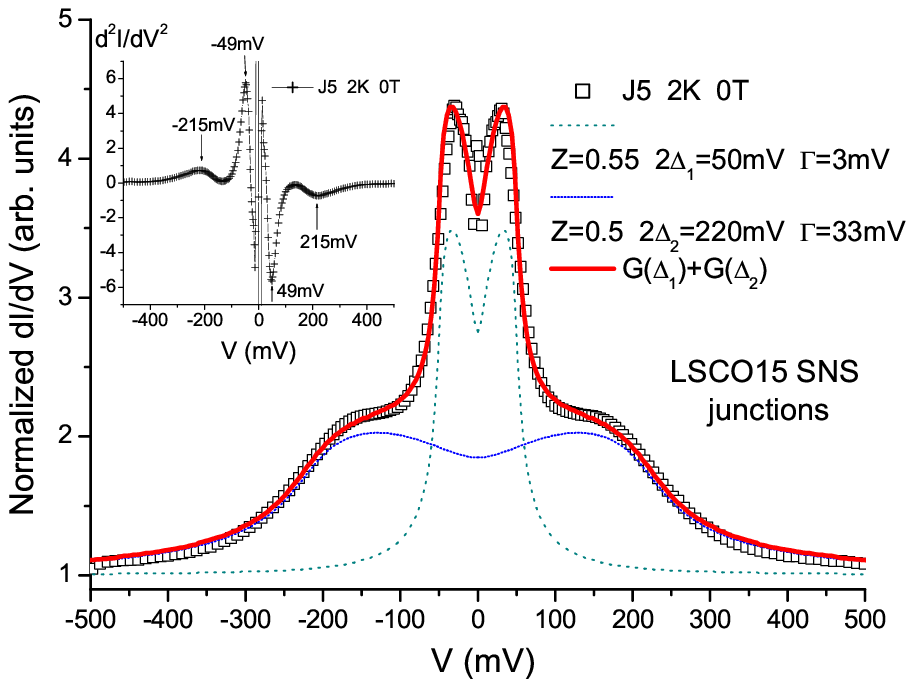}\vspace{5mm}

\noindent Fig. 4S:  Conductance spectrum of an anti-node S-N-S
junction of LSCO15-LSCO35-LSCO15 at 2K with a fit to the BTK model
for a d-wave superconductor. The two components of the fit with
$\Delta_1$ and $\Delta_2$ are also shown. The inset shows the
derivative of the conductance data of this junction.
\end{figure}

\section{Additional conductance spectra of LSCO15-LSCO35-LSCO15\\
S-N-S junctions}

In Figs. 4S, 5S and 6S we present normalized conductance spectra
of the LSCO15-LSCO35-LSCO15 S-N-S junctions, which were omitted
from the main paper for lack of space. These figures show three
spectra with the largest, smallest and intermediate $\Delta_2$
energy gap values, respectively. Note that this systematics of the
$\Delta_2$ values does not apply strictly to the $\Delta_1$
values. The $2\Delta_2$ feature in Fig. 4S is observed most
clearly due to the large separation between the $2\Delta_2$=220 mV
and $2\Delta_1$=50 mV values. This is also seen very clearly in
the second derivative data of the inset to this figure. We stress
here that in Fig. 5 of the main paper the $\Delta_2$ values are
mean values of all the working junctions on each wafer with their
statistical errors.  In the LSCO15 S-N-S junctions case there were
7 working junctions on the wafer, which yielded a mean
$\Delta_2(x=0.15)$ value of 76 mV and an error of $\pm 11$ mV. The
additional data of Figs. 4S, 5S and 6S show the robustness of the
second $2\Delta_2$ feature with 0.15\% Sr doping, which behaves on
the phase diagram of Fig. 5 as a clear peak rather than a flat
dome-like feature. We have no idea at the present time as for the
origin of this behavior, except for saying that this behavior is
apparently due to the optimal doping of these junctions. We can
however point out the similarity between the phase diagram results
of $\Delta_2$ in Fig. 5 at 2 K and the Nernst result of
$T_{onset}$ above $T_c$ (see Fig. 20 of Ref. \cite{Ong}).
Possibly, the current in our junctions decreases the phase
stiffness of the condensate, leading to an uncorrelated pairs
scenario similar to the one believed to occur above $T_c$. If this
is actually the case, the relevant energy or temperature scales
might be $\Delta_2$ here, or $T_{onset}$ in the Nernst case,
respectively, but not $T^*$ of the pseudogap.\\

\begin{figure} \hspace{-20mm}
\includegraphics[height=9cm,width=13cm]{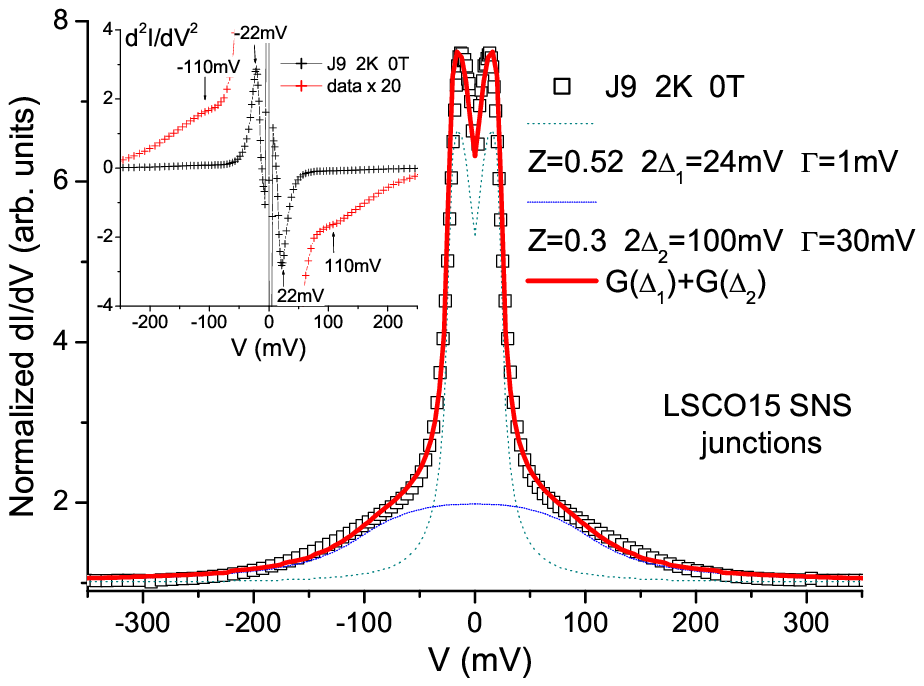}\vspace{5mm}

\noindent Fig. 5S:  Conductance spectrum of an anti-node S-N-S
junction of LSCO15-LSCO35-LSCO15 at 2K with a fit to the BTK model
for a d-wave superconductor. The two components of the fit with
$\Delta_1$ and $\Delta_2$ are also shown. The inset shows the
derivative of the conductance data of this junction.
\end{figure}

\begin{figure} \hspace{-20mm}
\includegraphics[height=9cm,width=13cm]{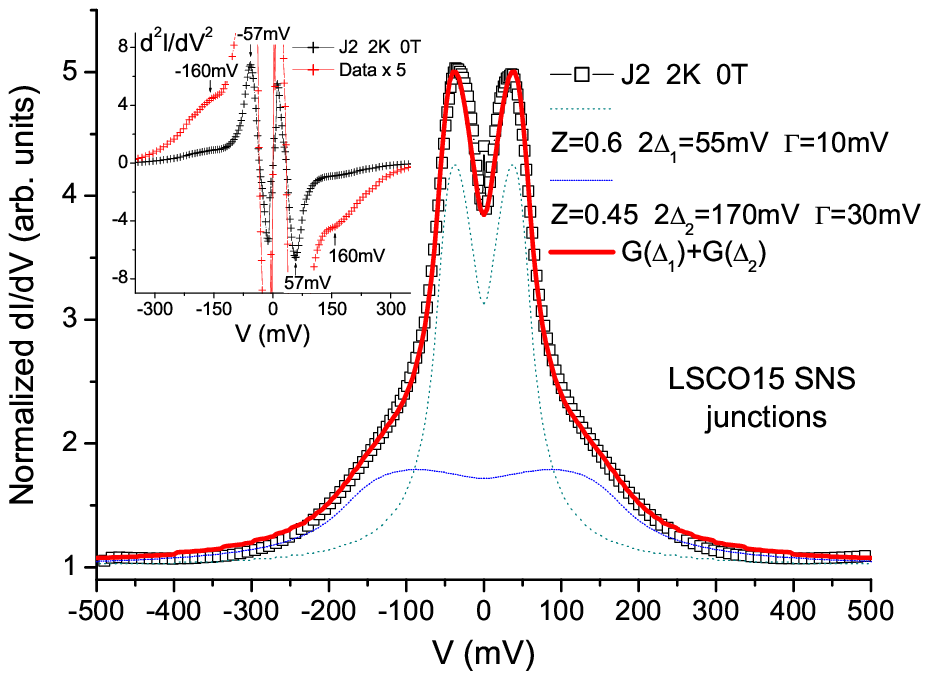}\vspace{3mm}

\noindent Fig. 6S:  Conductance spectrum of an anti-node S-N-S
junction of LSCO15-LSCO35-LSCO15 at 2K with a fit to the BTK model
for a d-wave superconductor. The two components of the fit with
$\Delta_1$ and $\Delta_2$ are also shown. The inset shows the
derivative of the conductance data of this junction.
\end{figure}

\section{High current density effects in the present study}

Another question that might arise concerns the high current
densities at high bias used in the present study in either the S-N
or S-N-S junctions. One might expect that nonlinear and heating
effects play a role and that the high measuring bias current takes
the junctions out of equilibrium. First, we estimate the highest
critical current densities at 2 K involved in this study. In the
the S-N-S junctions, the highest currents (at highest bias of 150
mV) range between 4 and 8 mA which correspond to current densities
of $1-2\times 10^6\, A/cm^2$. In the S-N junctions the highest
currents (at highest bias of 700 mV) are of about 5-6 mA with
corresponding current densities of less than $1.5\times 10^6\,
A/cm^2$. We note that the critical current density of the
superconducting electrodes at 2 K is of about $10\times 10^6\,
A/cm^2$. It is thus concluded that the highest current densities
in the superconducting electrodes are between 5 and 10 times
smaller than the critical current density and no nonlinear or
heating effects are expected. Such effects however, can still
occur in the junctions and the normal electrodes. We argue that
due to the very short relaxation time of the quasiparticles, on
the order of the inelastic scattering time in solids which is on
the order of $10^{-12}$ second, the system has time to relax to
its equilibrium state, certainly on the time scale of the
measurements which is on the order of 1 second (actually a DC
measurement). Furthermore, since we measured the correct
superconducting $\Delta_1$ values at $0.5-1\times 10^6\, A/cm^2$,
it is hard to believe that at $1-2\times 10^6\, A/cm^2$ where the
$\Delta_2$ feature was observed, a sudden change took the system
out of equilibrium. In any case, Andreev spectroscopy of highly
transparent junctions necessitates higher current densities than
tunneling spectroscopy, so that these high current density values
can not be avoided. Heating at high bias current can still be a
problem. We do see some heating effects occasionally, but it is
easy to detect them and stop the measurements in these cases. But
the more important fact is that measurements at temperatures
higher than 2 K, say at 4.3 or 6.6 K as in Fig. 4 of the main
paper, show very small changes of the conductance spectra.
Therefore, heating by 1-2 K at 2 K  will not affect our results at
all. We can thus conclude this section by saying that under the
present experimental conditions nonlinear and heating effects do
not play a major role.\\

\bibliography{AndDepBib.bib}

\bibliography{apssamp}

\end{document}